# Azobenzene photoisomerization probes cell membrane nanoviscosity

*Arianna Magni,[†,‡] Gaia Bondelli,[†,‡] Giuseppe M. Paternò,[†] Samim Sardar,[†] Valentina Sesti,[§,†] Cosimo D'Andrea,[†,‡] Chiara Bertarelli,[§,†] Guglielmo Lanzani*[†,‡]*

[†] Center for Nano Science and Technology, Istituto Italiano di Tecnologia, Via Pascoli 70, 20133, Milano, Italy

[‡] Dipartimento di Fisica, Politecnico di Milano, Piazza L. da Vinci 32, 20133 Milano, Italy

[§] Dipartimento di Chimica, Materiali e Ingegneria Chimica 'Giulio Natta', Politecnico di Milano, Piazza L. da Vinci 32, 20133 Milano, Italy

**Corresponding Author**

* Email: guglielmo.lanzani@iit.it

**ABSTRACT** The viscosity of cell membranes is a crucial parameter that affects the diffusion of small molecules both across and within the lipidic membrane and that is related to several diseases. Therefore, the possibility to measure quantitatively membrane viscosity on the nanoscale is of great interest. Here, we report a complete investigation of the photophysics of an amphiphilic membrane-targeted azobenzene (ZIAPIN2) and we validate its use as viscosity probe for cell membranes. We exploit ZIAPIN2 the *trans-cis* photoisomerization to develop a molecular viscometer and to assess the viscosity of *Escherichia coli* bacteria membranes employing time-resolved fluorescence spectroscopy. Lifetime measurements of ZIAPIN2 in *E. coli* bacteria suspensions correctly indicate that membrane viscosity decreases as the samples were heated up. Our results report a membrane viscosity value in live *E. coli* cells going from 10 to 5 cP, increasing the temperature from 22 °C up to 40 °C.

**TOC GRAPHICS**

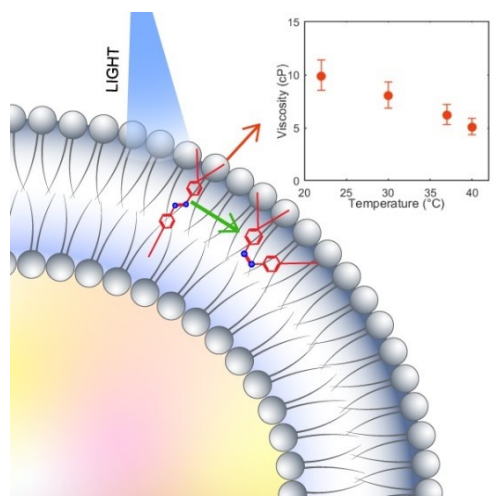

Cell membrane viscosity alteration is related to a large number of diseases, such as atherosclerosis, Alzheimer's, diabetes, and cell malignancy.[1,2] In general, the precise determination of viscosity at the molecular scale represents itself an important problem in cellular biophysics, as membrane viscosity influences many physiological processes, drug delivery and diffusion in cells and tissues.[3] Thus, measuring local viscosity in a micron or sub-micron region is an important challenge, as traditional mechanical methods (e.g. falling-ball viscometer, vibrational viscometers, rotational viscometers) used for probing viscosity are not suitable for living cells. Moreover, reliable probes at the molecular scale are essentially missing.

Several techniques have been exploited to estimate the viscosity of biological membranes. Most of them relies on spectroscopy and microscopy, so that one can infer viscosity by measuring some physical observables whose dependence on viscosity is known. One approach involves the measurement of the two-dimensional lateral diffusion coefficient, following the transport of fluorescent molecules. This can be done for instance exploiting Fluorescence Recovery after Photobleaching (FRAP),[4,5] Fluorescence Correlation Spectroscopy (FCS)[6,7] and single particle tracking.[8,9] Other methodologies include the use of fluorescent probes able to modify their emission properties according to their microenvironment. To this end, different molecular probes have been synthetized to target the plasma membrane and specific organelles or cell compartments. Moreover, various spectroscopic observables can be linked to the viscosity of the environment. In this context, measurements of ratiometric fluorescence,[2,10] time resolved fluorescence[11–14] and steady-state and time-resolved fluorescence anisotropy[15] have been reported as techniques for estimating the viscosity in living cells. In recent years, the spectroscopic characterization has been coupled to microscopy allowing viscosity-imaging.[2] Molecules known as molecular rotors are commonly used as viscosity probes, and the most common molecular rotors include BODIPY-based complexes, DCVJ and CCVJ.[16]. The fluorescence emission of these compounds is strongly affected by the viscosity of the environment. After light excitation, a molecular rotor undergoes radiative decay emitting photons and non-radiative relaxation via intramolecular torsion. The latter is modified by the environment viscosity so that the non-radiative decay rate is enhanced in fluid media. For this reason, both the PL quantum yield and the lifetime are susceptible to viscosity.

In this work, we report a detailed investigation of the photophysics of ZIAPIN2, a recently synthesized membrane-targeted azobenzene,[17–20] and we demonstrate and propose its use as a viscosity probe for the cell membrane. The class of compounds the ZIAPIN2 belongs to undergoes *trans → cis* isomerization upon illumination with visible radiation, with the reverse *cis → trans* isomerization driven either by light of a different wavelength or by thermal relaxation. ZIAPIN2, however, exhibits additional peculiarities, namely it is amphiphilic, and has a suitable size to match cell membrane thickness. Consequently, ZIAPIN2 possesses a non-covalent affinity for the plasma membrane. When added to cell culturing media, ZIAPIN2 locates into the cell membrane with relatively high efficiency (close to 80%) and long permanence (at least up to 100 h).[17,19]

Here, we focus our attention on the effect of the environment viscosity onto the molecular photophysics. We performed an accurate spectroscopic characterization of the molecule dissolved in dimethyl sulfoxide (DMSO, viscosity 2.4 cP [21]) and glycerol (viscosity 1400 cP [21]) mixtures, in order to understand how the viscosity of the environment affects the isomerization properties of this azobenzene derivative. We conclude that ZIAPIN2 can be a reliable probe of local viscosity and we demonstrate this by measuring the viscosity of the bacteria cell membrane.

The absorption spectrum of ZIAPIN2 in DMSO (Figure 1) exhibits a non-homogeneously broaden vibronic structure, peaking at 470 nm, corresponding to the second vibronic replica, suggesting the Huang-Ryss factor for the major vibrational progression to be larger than 1. Adding glycerol to the solution, *i.e.* increasing the viscosity of the environment, the spectrum looses its vibronic structure, apparently due to a further broadening of the vibronic replicas that merge into a Gaussian lineshape. Furthermore, going from a 100% DMSO to a 100% glycerol solution, ZIAPIN2 absorption peak shifts to the red by 20 nm. The emission spectrum (Figure 1A) undergoes a larger bathochromic shift, upon increasing of the glycerol fraction in the solution, passing from 525 nm to 590 nm. Therefore, the Stokes shift becomes larger in a more viscous solvent, going from 55 nm in DMSO to 100 nm in the glycerol solution. The bathochromic shift in emission is typical of fluorophores that aggregate when dissolved in specific solvents. However, we have already reported that when ZIAPIN2 aggregates in water,[18] it has an emission peak at 620 nm, that is at longer wavelengths than what observed for the molecule in pure glycerol. This argues against aggregation and points to an intramolecular mechanism. We conjecture that upon optical absorption the molecule undergoes a substantial geometrical relaxation before emission, and that the excited state $S_1$ is stabilized in less fluid media.[22] Furthermore, we observe that the photoluminescence (PL) intensity of the fluorophore increases markedly in presence of glycerol (Figure 1B), an effect which clearly indicates that viscosity hinders the isomerization reaction, favoring the radiative decay.

This last hypothesis is confirmed by inspection of the transient absorption measurement (Figure S1) and by the excitation-emission profiles shown in Figure 2A-C. We obtained such profiles recording the PL intensity at a fixed emission wavelength, selected in the range 500 - 600 nm, with 10 nm step, and scanning the excitation wavelength over the spectral band 300 – 550 nm. If the excitation profile overlaps with the absorption spectrum, the Vavilov-Kasha rule is fulfilled:[23] following absorption, the molecule relaxes to the lower excited state before emission takes place. In the *trans* isomer, however, the radiative deactivation competes with the isomerization path. The trajectory of the wave packet, initially placed in the Franck-Condon region of the excited potential energy surfaces (PES), branches into two paths, as depicted in Figure 2D.[24] The most likely is the diabatic path to the conical intersection connecting to the *cis* isomer ground state. The other one, with minor probability, reaches a minimum in the PES, from which radiative decay to the ground state (emission peaked at 580 nm) can take place. Upon photoexcitation, the *cis* isomer population builds up, as a consequence of the isomerization. The *cis* isomer in turn absorbs light and re-emits if the excitation reaches the second excited state $S_2$, (absorption at 370 nm). For absorption transitions occurring at longer wavelength, *i.e.* ending up into the $S_1$ state, the *cis* isomer follows the diabatic trajectory back to the conical intersection with $S_0$. Note that emission from $S_2$ in the *cis* isomer is an example of anti-Kasha behavior. Indeed, the second excited state $S_2$ has two local minima close to the *trans* and *cis* geometries. It has been shown that from $S_2$ the molecule can reach $S_1$, due to thermal relaxation, and then arrives on the ground state without emission of photons.[25,26] To highlight different relaxation paths, we superimposed the absorption spectra of the two isomers to the PL excitation profiles. We used the absorption spectrum of the *trans* isomer measured in DMSO (same as Figure 1), and the absorption spectrum of the *cis* isomer as taken from ref. 18. Interestingly, the excitation-emission profiles (Figure 2A-C) exhibit different features, depending on the viscosity of the environment. When ZIAPIN2 is dissolved in DMSO (Figure 2A), the excitation profile shows a maximum around 370 nm, corresponding to the absorption band of the *cis* conformation of the molecule, which has the anti-Kasha PL emission peak around 500 nm. According to the picture described above the absorption peak of the *trans* conformer is not clearly distinguishable due to the efficient isomerization, even if the *trans* population is predominant at room temperature. On the other hand, when ZIAPIN2 is dissolved in a high viscous medium (100% glycerol, Figure 2B), the excitation profile has a maximum around 480 nm, associated with PL peaked around 580 nm, and a second excitation maximum at 330 nm. Both these features are a fingerprint of the *trans* isomer absorption. We infer that viscosity hampers, or dramatically slows down, the pedal-like torsion of the azobenzene unit leading to isomerization. Upon absorption, the most likely trajectory of the wave packet is towards the emitting site. In intermediate conditions, that is when the molecule is dissolved in a mixture of DMSO and glycerol (75% DMSO, 25% glycerol, Figure 2C), both the *trans* and the *cis* isomer spectral features are present.

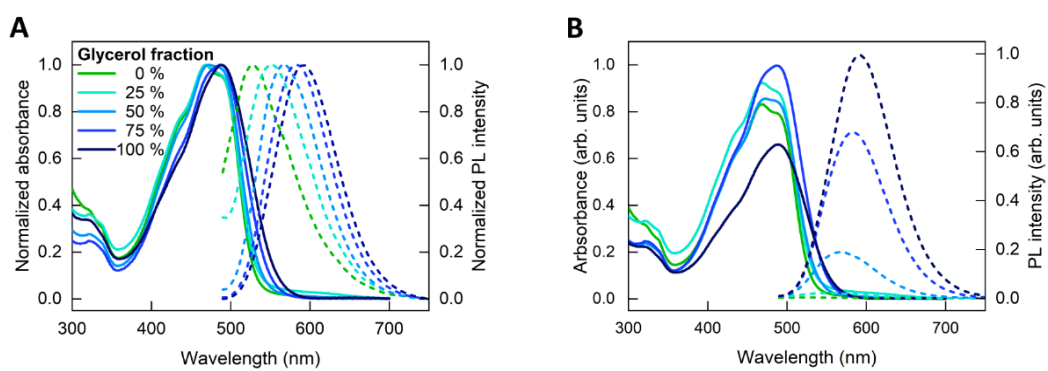

**Figure 1.** Absorption spectra (solid line) and emission spectra (dotted line) of ZIAPIN2 in DMSO and glycerol mixtures, normalized to the maximum (A) and with respect to the sample absorption around 470 nm (B). The emission spectra were acquired exciting the samples at 470 nm.

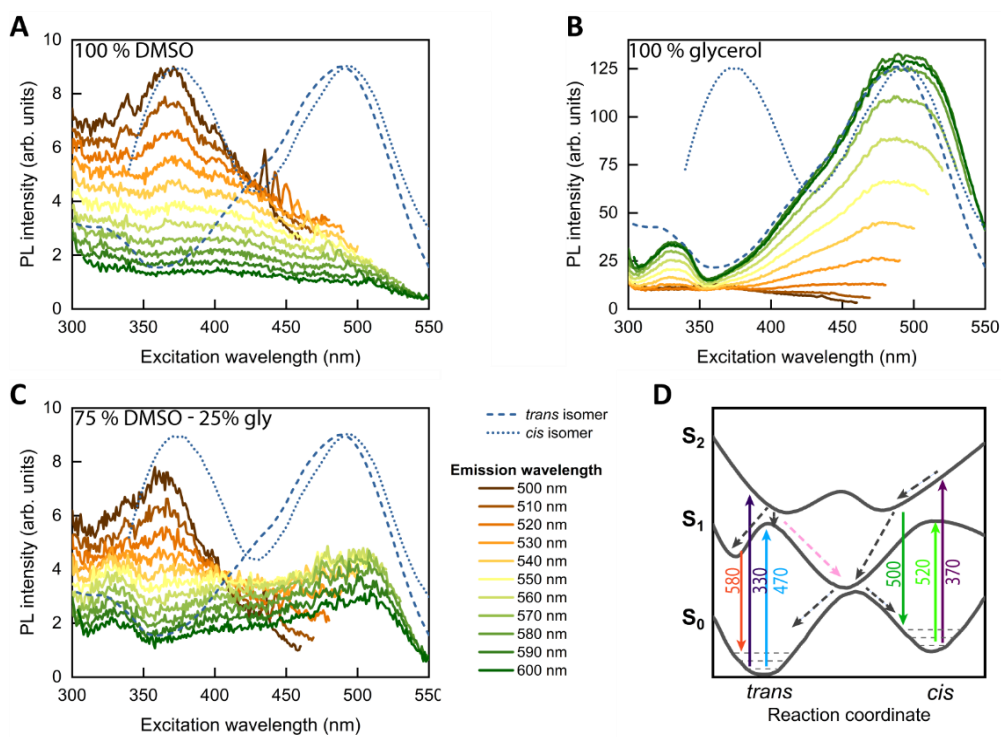

**Figure 2.** Excitation - emission profiles of ZIAPIN2 in (A) DMSO, (B) glycerol and (C) 75% DMSO – 25% glycerol mixture. For each curve in plots (A-C) the emission wavelength is fixed at a value between 500 and 600 nm, with 10 nm steps. The dotted lines represent the absorption spectra of the two isomers of the molecule, that have been superimposed to the excitation profiles. The scheme in (D) (adapted from [34]) shows the potential energy surfaces (PES) and an indication of the allowed optical transitions (solid arrows), while the dotted arrows represents the non-radiative transitions. Between $S_0$ and $S_1$ there is a conical intersection, responsible for the Kasha's rule violation.

The previous discussion suggests that ZIAPIN2 fluorescence, being strongly sensitive to the viscosity of the environment, is a good candidate for building a molecular viscometer. In addition, given its strong affinity for biological cell membranes, ZIAPIN2 can be a tool for estimating viscosity in lipid bilayers on the nanoscale. With this in mind, we selected the fluorescence lifetime as the spectroscopic observable that we use as viscosity indicator. To assess the validity of this choice, we performed time-resolved photoluminescence (TRPL) measurements on ZIAPIN2 dissolved in the same DMSO and glycerol mixtures. Indeed, we observed that in more viscous media the fluorescence decay becomes slower.[27] A quantitative analysis was performed in the spectral region between 550 and 590 nm (Figure 3A), by assuming a bi-exponential decay model. The fitting results are reported in Table S1 and they show that both time constants $\tau_1$ and $\tau_2$ increase with the viscosity of the environment. Moreover, the weight of the faster component ($A_1$) goes down with the increase of the glycerol fraction. The observed behavior is combined in the mean fluorescence lifetime $\tau_M$, which increases with viscosity. This reflects the role of friction onto isomerization, which is a process that occurs in a few picoseconds.

In order to construct the molecular viscometer, we built-up a calibration scale of viscosity *vs.* average lifetime $\tau_M$ (see Eq.(4)) by measuring the fluorescence lifetime of ZIAPIN2 in mixtures of known viscosity (DMSO and glycerol mixtures) in a range from 2.4 cP to 1460 cP.[21] The calibration has been achieved following these steps. Firstly, we estimated the viscosity of the mixture of DMSO and glycerol at room temperature (see Table S1 and Material and Methods). Secondly, we introduced a functional dependence of the viscosity $\eta$ onto the average fluorescence lifetime $\tau_M$ according to the power law suggested by the Förster-Hoffmann equation [28]

$$\log \tau_M = \alpha \log \eta + \log \frac{k_r}{z} \tag{1}$$

where $k_r$ is the radiative decay rate and $\alpha$ and $z$ are arbitrary constants.

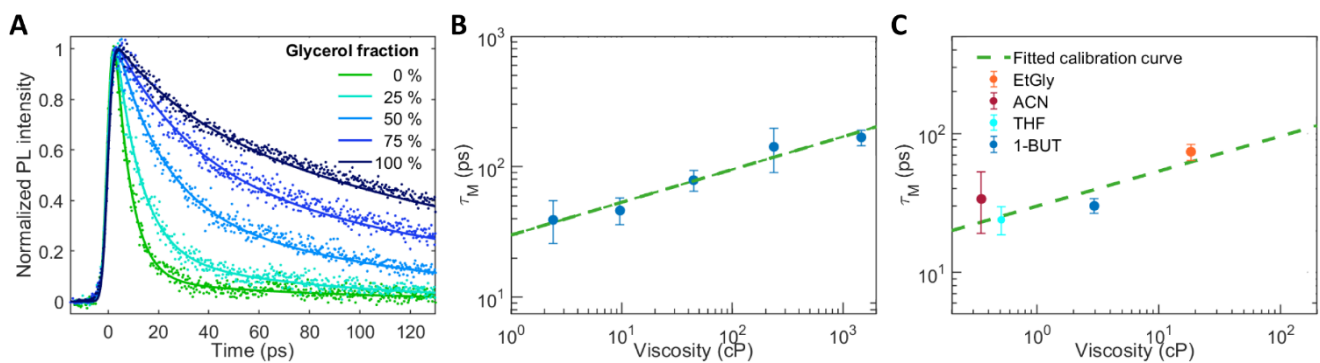

**Figure 3.** (A) PL dynamics of ZIAPIN2 in mixtures of DMSO and glycerol. The PL decays are plotted in the spectral region 550 - 590 nm. The scattered points represent the data while the solid lines the fitted curves. (B) Fitted average fluorescence lifetime of ZIAPIN2 in DMSO and glycerol mixtures as function of the viscosity of the solution. The blue points represent the five different samples. The error bars are calculated considering the 95% confidence interval given by the PL decays fitting. The central dashed line is the calibration curve, obtained as regression curve of the five points above. (C) The decays of ZIAPIN2 in different solvents have been analyzed to validate the viscosity-calibration with good agreement.

Figure 3B shows the average fluorescence lifetime as a function of the medium viscosity. The error bars are computed considering the 95% confidence interval on the fitted parameters and exploiting error propagation laws of Eq. (4). The experimental points are well aligned, and the calibration curve can be obtained using linear regression. The values found are $\alpha = 0.25 \pm 0.03$ and $\frac{k_r}{z} = 1.47 \pm 0.06$ (values $\pm$ standard error of the estimation), leading to

$$\eta = 10^{-5.87} \times \tau_M^{3.97} \tag{2}$$

To validate the calibration curve, we examined the fluorescence decay of ZIAPIN2 in other solvents of known viscosity. The chosen solvents were ethylene glycol (EtGly, 18.4 cP at RT),[21] and tetrahydrofuran (THF), 1-butanol (1-BUT), acetonitrile (ACN), exhibiting similar viscosities (0.51, 2.96 and 0.35 cP, respectively [21]) but different polarities ($\varepsilon_r$ 7.58, 17.84 and 36.64, respectively [21]). The results of the fittings are summarized in Table S2. Data, as reported in Figure 3C, show good agreement with our calibration scale. In particular, the polarity of the solvents does not have a strong influence on the fluorescence dynamics, corroborating the idea to adopt ZIAPIN2 as a viscosity probe.

Finally, we tested our molecular viscometer in the case study of *Escherichia coli* (*E. coli*) cell membrane. ZIAPIN2 preferentially localizes into the cell membrane,[17,18] thus we can neglect the contribution of the small fraction (below 20%) of molecules that eventually remain both inside and outside the bacterial cells. Furthermore, we considered the fluorescence decays in the spectral region between 550 and 590 nm, filtering out the PL peak of ZIAPIN2 in water, that is around 620 nm (Figures S2-3).[18] The fluorescence decay curves of ZIAPIN2 in *E. coli* at various temperatures are shown in Figure 4A, together with the fitted curves. The fitted parameters are reported in Table S3. Viscosity, as a function of temperature, is plotted in Figure 4B. The error bars are computed evaluating equation (5) at the edges of the 50% confidence interval for $\tau_M$. Despite the small variations in the decay kinetics, our viscometer allows to extract useful information. We noticed that increasing the sample temperature, the fluorescence decay becomes faster, indicating that the ZIAPIN2 *trans - cis* isomerization reaction is favored, and that the membrane becomes more fluid, in agreement with the existing literature.[29] In the selected temperature range, the viscosity changes from 10 cP to 5 cP, denoting a reduction of *E. coli* membrane viscosity of around 50 % going from room temperature to 40 °C.

At present, all the attempts made for estimating the viscosity of cell membranes exploited indirect measurement techniques that are intrinsically problematic, and no ultimate protocol has been proposed so far. We decided to exploit ZIAPIN2 as viscosity probe, not only because of its sensitivity to the environment, but also for its ability of targeting cell membranes and for its low toxicity.[17] To the best of our knowledge, there is only one other work reporting on viscosity of *E. coli* membrane.[11] In that work, however, the authors reported a viscosity of 950 cP, exceeding by almost two order of magnitude our estimate. Such a discrepancy might originate from several factors. Firstly, it has been shown that BODIPY-based probes, as the one used in 11, have PL lifetimes that are strongly dependent on the polarity of the environment.[30,31] Moreover, the membrane consists of large macromolecules, which may lead to the underestimation of the bulk viscosity of the system. Indeed, all molecular rotors are only able to sense the local viscosity, in a volume comparable to the probe size.[30] Furthermore, ZIAPIN2 has an average PL lifetime that ranges from tens to hundreds of picoseconds, which is a much shorter timescale (around two orders of magnitude) compared to other commercially available BODIPY-based molecular rotors. This suggests different mechanisms of conformational adjustment in the two molecules. Finally, it is worth reminding that the large biological variability could lead to a large uncertainty in the results. All these factors indicate the lack of reliable tools for measuring membrane viscosity and strengthen the value of our work, which aims at filling the gap in the present technology.

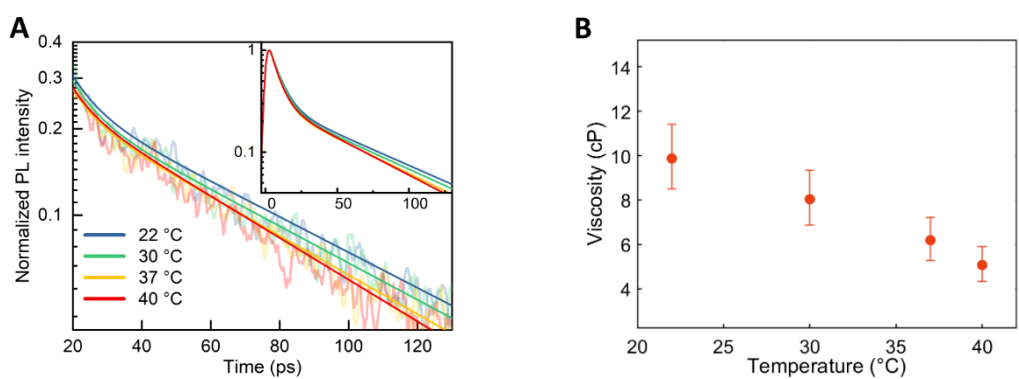

**Figure 4.** (A) PL dynamics of ZIAPIN2 in *E. Coli* suspensions at different temperatures. The PL decays are plotted in the spectral region 550 - 590 nm. The plot shows both the data and the fitted curves. In the insect the fitted curves are shown in the whole acquisition time window. (B) Estimated viscosity of *E. coli* membranes at different temperatures. The points are the ones achieved using the ZIAPIN2 – viscometer. The error bars were obtained as explained in the text.

In conclusion, we studied the photophysical properties of a recently synthesized azobenzene-based molecule (ZIAPIN2) showing that the isomerization mechanism is strongly dependent on the viscosity of the environment. In high viscous media, the isomerization is hindered, an effect that leads to an increase of the PL quantum yield and of the excited state lifetime. Given the amphiphilic nature of ZIAPIN2, we exploited these features to build a molecular viscosity sensor to probe the *E. coli* membrane. We quantitatively estimated the viscosity of the bacterial membrane at different temperatures. We thus deem ZIAPIN2 fluorescence properties to be a new valuable tool for nanoviscosity measurements in lipid membranes.

**Materials and methods**

**Steady-state UV-Vis and PL measurements**  UV-Vis absorption measurements were performed using a Perkin ElmerLambda 1050 spectrophotometer, with deuterium (180–320 nm) and tungsten (320–3300 nm) lamps, a monochromator and three detectors (photomultiplier 180–860 nm, InGaAs 860–1300 nm, and PbS 1300–3300 nm). Absorption spectra were normalized according to a reference spectrum taken at 100% transmission (without the sample), 0% transmission (with an internal shutter), and in the presence of the reference solvent.

For the PL measurements and the excitation profiles an iHR320Horiba NanoLog Fluorometer was employed, equipped with a Xenon lamp, two monochromators, and two detectors (photomultiplier and InGaAs). Samples were excited at 470 nm.

**Time-resolved photoluminescence (TRPL) measurements**  TRPL measurements were carried out using a femtosecond laser source coupled to a streak camera detection system (Hamamatsu C5680). A Ti:sapphire laser (Coherent Chameleon Ultra II, pulse bandwidths of ∼140 fs, repetition rate of 80 MHz, and maximum pulse energy of 50 nJ) was used to pump a second-harmonic crystal (β-barium borate) to tune the pump wavelength to 470 nm. The measurements here shown were performed recording the first 130 ps of decays, with an IRF of 4.1 ps.

When required, a Peltier cell was used in order to control the samples temperature.

**Analysis**  The analysis of the TRPL decay kinetics was performed in the spectral re, gion between 550 and 590 nm (Figure 3A), by assuming a biexponential decay model

$$I(t) = A_1 \exp\left(-\frac{t}{\tau_1}\right) + A_2 \exp\left(-\frac{t}{\tau_2}\right) \tag{3}$$

where $A_1$, $A_2$, $\tau_1$, and $\tau_2$ are the amplitudes and lifetimes of the two exponentially decaying components, respectively. The curves have been fitted considering also the convolution of the biexponential decay with the instrument response function (IRF), that is a Gaussian with FWHM of 4.1 ps.

The observable of interest is the mean fluorescence lifetime $\tau_M$ calculated as follows

$$\tau_M = \frac{A_1 \tau_1^2 + A_2 \tau_2^2}{A_1 \tau_1 + A_2 \tau_2} \tag{4}$$

To estimate the viscosity of the mixture of DMSO and glycerol at room temperature, we exploited the Nissan-Grunberg relation with experimental parameters taken from literature.[32,33] We used the following expression to calculate the viscosity $\eta_{mix}$ of the DMSO and glycerol mixtures, as a function of the molar fraction $x$ of the two components.

$$\eta_{mix} = \exp(x_{GLY} \ln \eta_{GLY} + x_{DMSO} \ln \eta_{DMSO} - 0.961 x_{GLY} \cdot x_{DMSO}) \qquad (5)$$

**Sample preparation** ZIAPIN2 was synthesized according to the procedure reported in ref. 17, and characterized by $^1$H-NMR in DMSO using a Bruker ARX400. The solutions were prepared by suspending the proper amount of ZIAPIN2 in the solvents in order to obtain a final concentration of 25 μM. The solvents used are DMSO and glycerol mixtures (0, 25%, 50%, 75%, 100% v/v of glycerol), ethylene glycol, acetonitrile, 1- butanol and tetrahydrofuran. All chemicals were purchased from Sigma Aldrich.

***E. coli* cultures** Experiments were conducted using *Escherichia coli* (*E. coli*) strain ATCC25922. For bacterial growth, a single colony was inoculated in Luria-Bertani (LB) broth and incubated overnight at 37 °C with shaking at 215 rpm, until stationary phase was reached. Then, bacterial suspension turbidity (expressed as optical density at 600nm; $OD_{600}$) was diluted to $OD_{600}$ = 1 in LB broth, without antibiotics. Bacteria were then centrifuged, and the obtained pellet was resuspended in a phosphate-buffered saline (PBS) aqueous solution. ZIAPIN2 was mixed with the bacterial suspension to obtain a final concentration of 25 μM. After 1 h of incubation, bacteria were again centrifuged and resuspended in fresh PBS, in order to remove the cell-unbounded molecules.

**Associated Content**

Supporting Information containing transient absorption data, PL spectra and dynamics of ZIAPIN2 in PBS water and the tables with the fitting results.

**Acknowledgments**

G.M.P thanks Fondazione Cariplo (Grant No. 2018-0979) for financial support.

# References

1. Haidekker, M. A. *et al.* New fluorescent probes for the measurement of cell membrane viscosity. *Chem. Biol.* **8**, 123–131 (2001).

2. Kuimova, M. K. Mapping viscosity in cells using molecular rotors. *Phys. Chem. Chem. Phys.* **14**, 12671 (2012).

3. Suhling, K. Twist and Probe—Fluorescent Molecular Rotors Image Escherichia coli Cell Membrane Viscosity. *Biophys. J.* **111**, 1337–1338 (2016).

4. Axelrod, D., Koppel, D. E., Schlessinger, J., Elson, E. & Webb, W. W. Mobility measurement by analysis of fluorescence photobleaching recovery kinetics. *Biophys. J.* **16**, 1055–1069 (1976).

5. Dayel, M. J., Hom, E. F. Y. & Verkman, A. S. Diffusion of Green Fluorescent Protein in the Aqueous-Phase Lumen of Endoplasmic Reticulum. *Biophys. J.* **76**, 2843–2851 (1999).

6. Wawrezinieck, L., Rigneault, H., Marguet, D. & Lenne, P.-F. Fluorescence Correlation Spectroscopy Diffusion Laws to Probe the Submicron Cell Membrane Organization. *Biophys. J.* **89**, 4029–4042 (2005).

7. Kim, S. A., Heinze, K. G. & Schwille, P. Fluorescence correlation spectroscopy in living cells. *Nat. Methods* **4**, 963–973 (2007).

8. Saxton, M. J. & Jacobson, K. SINGLE-PARTICLE TRACKING:Applications to Membrane Dynamics. *Annu. Rev. Biophys. Biomol. Struct.* **26**, 373–399 (1997).

9. Manzo, C. & Garcia-Parajo, M. F. A review of progress in single particle tracking: from methods to biophysical insights. *Reports Prog. Phys.* **78**, 124601 (2015).

10. Zhao, M. *et al.* A water-soluble two-photon fluorescence chemosensor for ratiometric imaging of mitochondrial viscosity in living cells. *J. Mater. Chem. B* **4**, 5907–5912 (2016).

11. Mika, J. T. *et al.* Measuring the Viscosity of the Escherichia coli Plasma Membrane Using Molecular Rotors. *Biophys. J.* **111**, 1528–1540 (2016).

12. Levitt, J. A. *et al.* Membrane-Bound Molecular Rotors Measure Viscosity in Live Cells via Fluorescence Lifetime Imaging. *J. Phys. Chem. C* **113**, 11634–11642 (2009).

13. Steinmark, I. E. *et al.* Targeted fluorescence lifetime probes reveal responsive organelle viscosity and membrane fluidity. *PLoS One* **14**, e0211165 (2019).

14. López-Duarte, I., Vu, T. T., Izquierdo, M. A., Bull, J. A. & Kuimova, M. K. A molecular rotor for measuring viscosity in plasma membranes of live cells. *Chem. Commun.* **50**, 5282–5284 (2014).

15. Levitt, J. A. *et al.* Fluorescence Anisotropy of Molecular Rotors. *ChemPhysChem* **12**, 662–672 (2011).

16. Lee, S.-C. *et al.* Fluorescent Molecular Rotors for Viscosity Sensors. *Chem. - A Eur. J.* **24**, 13706–13718 (2018).

17. DiFrancesco, M. L. *et al.* Neuronal firing modulation by a membrane-targeted photoswitch. *Nat. Nanotechnol.* **15**, 296–306 (2020).

18. Paternò, G. M. *et al.* Membrane Environment Enables Ultrafast Isomerization of Amphiphilic Azobenzene. *Adv. Sci.* **7**, 1903241 (2020).

19. Vurro, V. *et al.* Molecular Design of Amphiphilic Plasma Membrane-Targeted Azobenzenes for Nongenetic Optical Stimulation. *Front. Mater.* **7**, 1–10 (2021).

20. Paternò, G. M. *et al.* The Effect of an Intramembrane Light-Actuator on the Dynamics of Phospholipids in Model Membranes and Intact Cells. *Langmuir* **36**, 11517–11527 (2020).

21. Cardarelli, F. *Materials Handbook*. (Springer International Publishing, 2018).

22. Lakowicz, Joseph R, E. *Principles of fluorescence Spectroscopy*. *Principles of fluorescence Spectroscopy* (1999).

23. Kasha, M. Characterization of electronic transitions in complex molecules. *Discuss. Faraday Soc.* **9**, 14 (1950).

24. Cembran, A., Bernardi, F., Garavelli, M., Gagliardi, L. & Orlandi, G. On the Mechanism of the cis-trans Isomerization in the Lowest Electronic States of Azobenzene: S0, S1, and T1. *J. Am. Chem. Soc.* **126**, 3234–3243 (2004).

25. Quick, M. *et al.* Photoisomerization dynamics and pathways of trans - And cis -azobenzene in solution from broadband femtosecond spectroscopies and calculations. *J. Phys. Chem. B* **118**, 8756–8771 (2014).

26. Rau, H. *Photoisomerization of Azobenzenes*. *Photoreactive Organic Thin Films* **1**, (Woodhead Publishing Limited, 2002).

27. Chang, C. W., Lu, Y. C., Wang, T. Te & Diau, E. W. G. Photoisomerization dynamics of azobenzene in solution with S1 excitation: A femtosecond fluorescence anisotropy study. *J. Am. Chem. Soc.* **126**, 10109–10118 (2004).

28. Förster, T. & Hoffmann, G. Viscosity dependence of fluorescent quantum yields of some dye systems. *Zeitschrift fur Phys. chemie* **75**, 63–76 (1971).

29. de Almeida, R. F. M., Fedorov, A. & Prieto, M. Sphingomyelin/Phosphatidylcholine/Cholesterol Phase Diagram: Boundaries and Composition of Lipid Rafts. *Biophys. J.* **85**, 2406–2416 (2003).

30. Polita, A., Toliautas, S., Žvirblis, R. & Vyšniauskas, A. The effect of solvent polarity and macromolecular crowding on the viscosity sensitivity of a molecular rotor BODIPY-C 10. *Phys. Chem. Chem. Phys.* **22**, 8296–8303 (2020).

31. Vyšniauskas, A. *et al.* Exploring viscosity, polarity and temperature sensitivity of BODIPY-based molecular rotors. *Phys. Chem. Chem. Phys.* **19**, 25252–25259 (2017).

32. Grunberg, L. & Nissan, A. H. Mixture Law for Viscosity. *Nature* **164**, 799–800 (1949).

33. Angulo, G. *et al.* Characterization of Dimethylsulfoxide / Glycerol Mixtures: A Binary Solvent System for the Study of "Friction-Dependent" Chemical Reactivity. *Phys. Chem. Chem. Phys.* **18**, 18460–18469 (2016).

34. Wang, Y., Li, C., Zhang, B., Qin, C. & Zhang, S. Ultrafast investigation of excited-state dynamics in trans -4-methoxyazobenzene studied by femtosecond transient absorption spectroscopy. *Chinese J. Chem. Phys.* **31**, 749–755 (2018).

# Supporting information

# Azobenzene photoisomerization probes cell membrane nanoviscosity

*Arianna Magni,[†,‡] Gaia Bondelli[†,‡], Giuseppe M. Paternò,[†] Samim Sardar,[†] Valentina Sesti,[§,†] Cosimo D'Andrea,[†,‡] Chiara Bertarelli,[§,†] Guglielmo Lanzani*[†,‡]*

[†] Center for Nano Science and Technology, Istituto Italiano di Tecnologia, Via Pascoli 70, 20133, Milano, Italy

[‡] Dipartimento di Fisica, Politecnico di Milano, Piazza L. da Vinci 32, 20133 Milano, Italy

[§] Dipartimento di Chimica, Materiali e Ingegneria Chimica 'Giulio Natta', Politecnico di Milano, Piazza L. da Vinci 32, 20133 Milano, Italy

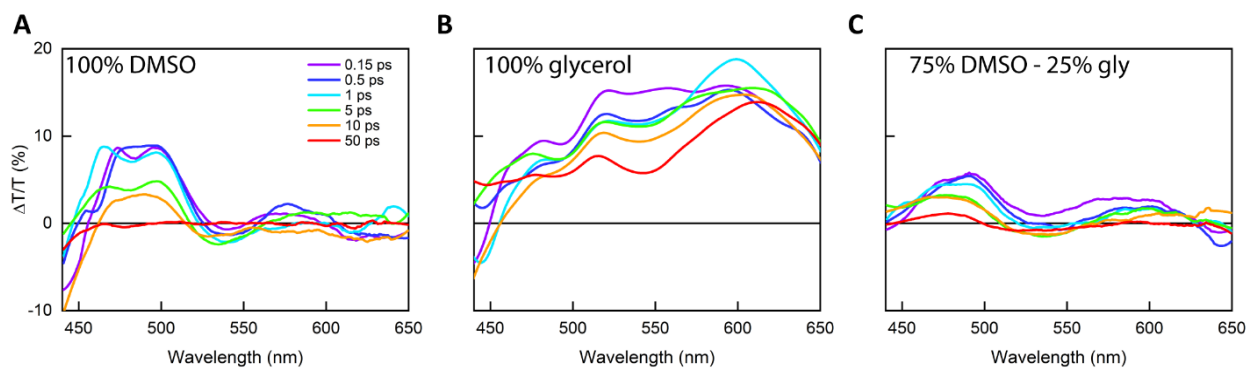

**Figure S1.** Differential transient transmittance spectra of ZIAPIN2 in (A) DMSO, (B) glycerol and (C) a mixture of the two (75% DMSO, 25% glycerol). Observing the spectra of panel (A) we can distinguish three different regions. Specifically, the spectral band between 460 nm and 500 nm shows a positive ΔT/T, due to the ground state bleaching (GSB) of the *trans* isomer of the molecule: after the excitation, part of the molecules is temporary stored in the excited state while others undergo isomerization. The spectral region between 520 nm and 560 nm exhibits a negative differential transmission as it corresponds to the *cis* form absorption band of ZIAPIN2. Finally, the 560–620 nm region has positive ΔT/T. This last spectral region corresponds to the PL peak of the fluorophore and hence positive values of differential transmission are attributed to stimulated emission (SE). The transient absorption spectra of ZIAPIN2 in glycerol (B) is larger than zero in the whole spectral range under investigation. In this case the signal has a different shape from the previous one, with the peak of the curves shifted around 600 nm, which represents the SE band peak of the molecule in a viscous media. The GSB signal is reduced with respect to the SE due to the absence of a long-living *cis* isomer and a favored radiative relaxation. In the case of intermediate viscosity (C), the SE band displays an increase in amplitude while the band corresponding to the *cis* isomer absorption decreases in amplitude, due to the reduced isomerization quantum yield with respect to (A).

These measurements were performed using a Ti:Sapphire laser with 2 mJ output energy, 1 kHz repetition rate, a pulse width of 100 fs and a central wavelength of 800 nm. Samples were pumped with 490 nm light generated with a visible optical parameter amplifier (OPA). Pump pulses were focused on a 200 μm spot (diameter), keeping a power of 100 μW. The white-light probe pulse was generated with a sapphire plate. The sample was contained in a quartz cuvette, so that the optical path length was 1 mm. The transmitted probe signal was collected by an optical multichannel amplifier (OMA).

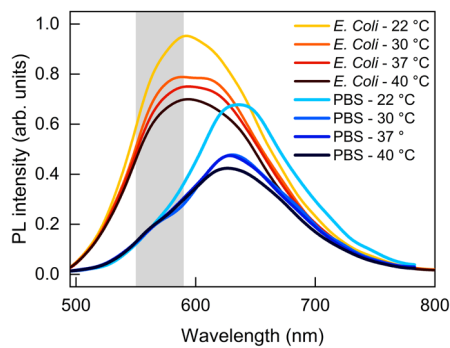

**Figure S2.** Fluorescence spectra of ZIAPIN2 in *E. Coli* suspensions and PBS at different temperatures, in the first 130 ps of the PL decay. As ZIAPIN2 fluorescence is strongly influenced by the medium, there is a significant spectral shift. In a water-like environment the PL peak is moved to 620 nm due to the formation of aggregates.[1] The huge spectral shift indicates that the molecule has a strong affinity for the bacteria membrane. Moreover, the fluorescence quantum yield decreases with temperature, being isomerization favored. The gray rectangle highlights the spectral region (550-590 nm) where the PL dynamics have been considered for our viscometer. Here the PL contribution of ZIAPIN2 molecules in water is minor.

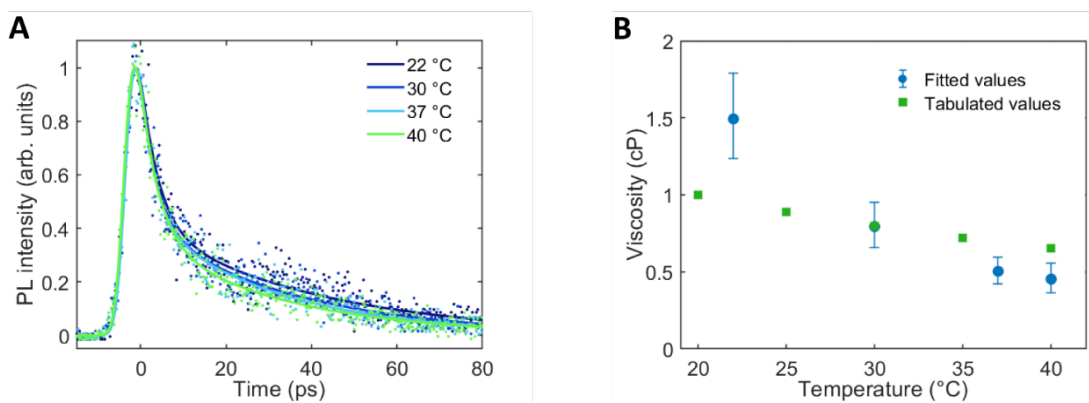

**Figure S3.** (A) PL dynamics of ZIAPIN2 in PBS buffer at different temperatures. The PL decays are in the region 550 - 590 nm. (B) Estimated viscosity of PBS buffer at different temperatures and tabulated values. The viscosity estimation is achieved using the ZIAPIN2 – viscometer. The error bars were obtained as reported in the main text.

**Table S1.** Fitting parameters of TRPL decay dynamics of ZIAPIN2 in DMSO and glycerol mixtures in the spectral range 550-590 nm. The decay kinetics have been fitted with a biexponential model, whose two components amplitudes and lifetimes are here reported. The mean fluorescence lifetime was calculated as in equation (4). The second column reports the viscosity calculated as in equation (5).

|  | Viscosity (cP) | $\tau_1$ (ps) | $A_1$ (%) | $\tau_2$ (ps) | $A_2$ (%) | $\tau_M$ (ps) | $R^2$ |
|---|---|---|---|---|---|---|---|
| **Glycerol 0%** | 2.4 | 6 | 94.1 | 81 | 5.9 | 39 | 0.984 |
| **Glycerol 25%** | 9.66 | 10 | 84.9 | 74 | 15.1 | 46 | 0.985 |
| **Glycerol 50%** | 44.5 | 18 | 63.4 | 98 | 36.6 | 79 | 0.987 |
| **Glycerol 75%** | 237 | 30 | 51.3 | 164 | 48.7 | 142 | 0.984 |
| **Glycerol 100%** | 1460 | 28 | 26.6 | 176 | 73.4 | 168 | 0.993 |

**Table S2.** Fitting parameters of TRPL decay dynamics of ZIAPIN2 in four different solvents in the spectral range 550-590 nm. The decay kinetics have been fitted with a biexponential model, whose two components amplitudes and lifetimes are here reported. The mean fluorescence lifetime was calculated as in equation (4).

|  | $\tau_1$ (ps) | $A_1$ (%) | $\tau_2$ (ps) | $A_2$ (%) | $\tau_M$ (ps) | $R^2$ | Tabulated viscosity[2] (cP) |
|---|---|---|---|---|---|---|---|
| **Ethylene glycol** | 16 | 58.8 | 89 | 41.2 | 74 | 0.988 | 18.3 |
| **Acetonitrile** | 3 | 98.4 | 90 | 1.6 | 34 | 0.991 | 0.35 |
| **THF** | 5 | 93.2 | 50 | 6.8 | 24 | 0.990 | 0.52 |
| **1-butanol** | 7 | 83.7 | 47 | 16.3 | 30 | 0.993 | 2.96 |

**Table S3.** Fitting parameters of TRPL decay dynamics of ZIAPIN2 in *E .Coli* suspension at different temperatures in the spectral range 550-590 nm. The decay kinetics have been fitted with a biexponential model, whose two components amplitudes and lifetimes are here reported. The mean fluorescence lifetime is calculated as in equation (4). The viscosity has been estimated according to the viscosity calibration curve hereby proposed.

| Temperature (°C) | $\tau_1$ (ps) | $A_1$ (%) | $\tau_2$ (ps) | $A_2$ (%) | $\tau_M$ (ps) | $R^2$ | Viscosity (cP) |
|---|---|---|---|---|---|---|---|
| **22** | 7 | 78.2 | 70 | 21.8 | 53 | 0.986 | 9.9 |
| **30** | 7 | 79.1 | 68 | 20.9 | 51 | 0.986 | 8.0 |
| **37** | 7 | 79.6 | 64 | 20.4 | 47 | 0.985 | 6.2 |
| **40** | 6 | 78.3 | 60 | 21.7 | 45 | 0.983 | 5.1 |

# References

1. Paternò, G. M. *et al.* Membrane Environment Enables Ultrafast Isomerization of Amphiphilic Azobenzene. *Adv. Sci.* **7**, 1903241 (2020).

2. Cardarelli, F. *Materials Handbook*. (Springer International Publishing, 2018).